\documentclass[11pt]{article}
\usepackage{amssymb}
\usepackage{graphics}
\usepackage{epsfig}
\usepackage{a4wide}
\usepackage{color}

\begin{document}

\newcommand{\eezzz}{$e^+ e^- \to Z^0 Z^0 Z^0~$}

\title{ Triple $Z^0$-boson production in large extra dimensions model at ILC }
\author{ \textcolor{red}{}
Jiang Ruo-Cheng, Li Xiao-Zhou, Ma Wen-Gan, Guo Lei, and Zhang Ren-You \\
{\small Department of Modern Physics, University of Science and Technology}\\
{\small of China (USTC), Hefei, Anhui 230026, P.R.China}  }

\date{}
\maketitle \vskip 15mm
\begin{abstract}
We investigate the effects induced by the interactions of the
Kaluza-Klein (KK) graviton with the standard model (SM) particles on
the triple $Z^0$-boson production process at the ILC in the
framework of the large extra dimension (LED) model. We present the
dependence of the integrated cross sections on the electron-positron
colliding energy $\sqrt{s}$, and various kinematic distributions of
final $Z^0$ bosons and their subsequential decay products in both
the SM and the LED model. We also provide the relationship between
the integrated cross section and the fundamental scale $M_S$ by
taking the number of the extra dimensions ($d$) as $3$, $4$, $5$,
and $6$, respectively. The numerical results show that the LED
effect can induce a observable relative discrepancy for the
integrated cross section ($\delta_{LED}$), which can reach the value
of $13.11\%~(9.27\%)$ when $M_S = 3.5~(3.8)~TeV$ and the colliding
energy $\sqrt{s} = 1~TeV$. We find the relative discrepancy of LED
effect can even reach few dozen percent in the high transverse
momentum area or the central rapidity region of the final
$Z^0$-bosons and muons.
\end{abstract}

\vskip 3cm {\large\bf PACS: 11.10.Kk, 13.66.Fg, 14.70.Hp}

\vfill \eject

\baselineskip=0.32in

\renewcommand{\theequation}{\arabic{section}.\arabic{equation}}
\renewcommand{\thesection}{\Roman{section}.}
\newcommand{\nb}{\nonumber}

%slash:
\newcommand{\Dir}{\kern -6.4pt\Big{/}}%su lettere italiane minuscole
\newcommand{\Dirin}{\kern -10.4pt\Big{/}\kern 4.4pt}
\newcommand{\DDir}{\kern -7.6pt\Big{/}}%su lettere italiane maiuscole
\newcommand{\DGir}{\kern -6.0pt\Big{/}}%su lettere greche

\makeatletter      % '@' is now a normal "letter" for TeX
\@addtoreset{equation}{section}
\makeatother       % '@' is restored as a "non-letter" character for TeX

\vskip 5mm
\section{Introduction}
\par
The large extra dimensions (LED) model, proposed by Arkani-Hamed,
Dimopoulos and Dvali (ADD) \cite{1}, is an attractive extension of
the standard model (SM) because of its possible testable
consequences. In the LED model only graviton can propagate in a
$D=4+d$ dimensional space with $d$ being the dimension number of
extra space, while the SM particles exist in the usual
$(3+1)$-dimensional space. The picture of a massless graviton
propagating in $D$-dimensions is equal to the scene where numerous
massive Kaluza-Klein (KK) gravitons propagate in $(3+1)$-dimensions.
So we can expect that even though the gravitational interactions in
the 4 space-time dimensions are suppressed by a factor of $1/M_P$,
they can be compensated by these numerous KK states. Therefore, in
either the case of real graviton emission or the case of virtual
graviton exchange, it is shown that the Plank mass $M_P$ cancels out
of the cross section after summing over the KK states, and we can
obtain an interaction strength comparable to the electroweak
strength \cite{2,3}. Up to now, many works on the LED phenomenology
at colliders have been done, including vector boson pair productions
and association productions of vector boson with graviton
\cite{4,5,6}.

\par
The precision measurements of the trilinear gauge boson couplings
(TGCs) are helpful for verification of non-Abelian gauge structure,
and the investigation of the quartic gauge boson couplings (QGCs)
can either confirm the symmetry breaking mechanism or present a
direct test on the new physics beyond the SM. The vector boson pair
production processes were extensively studied in the SM for probing
the TGCs \cite{8,9,10,11,12}. The direct study of QGCs requires
the investigation of the processes involving at least three external
gauge bosons. Recently, the triple $Z^0$-boson production in the
LED model at the LHC was studied in Ref.\cite{7}.

\par
The International Linear Collider (ILC) is proposed with the
colliding energy $\sqrt{s} = 200 \sim 500~GeV$ which would be
upgraded to $\sqrt{s} = 1~TeV$, and the integrated luminosity is
required to be $1000~fb^{-1}$ in the first phase of operation
\cite{12}. The triple gauge boson productions at the ILC are
important processes in probing the TGCs and QGCs of electroweak
theory, which are related to electroweak symmetry breaking
mechanism. If their observables coincide with the LED predictions
on the triple gauge boson production processes, it means the
coupling of graviton to gauge bosons in the LED model would be
the causation. Therefore, the understanding of the LED phenomenology
in triple gauge boson production processes at the ILC is necessary.

\par
In this paper we study the LED effects on the process \eezzz at the
ILC. The paper is organized as follows: In section II we present the
calculation descriptions for the process with a brief review of the
related LED theory. The numerical results and discussions are given
in section III. In the last section a short summary is given.

\vskip 5mm
\section{Analytic calculations}
\par
In the LED model, the extra dimensions on a torus are compactified
to a radius $R/2 \pi$. The relationship between the usual Planck
scale $M_P$ to the fundamental scale $M_S$ is
\begin{equation}
M_P^2 = {2(4\pi)^{-{d/2}}\over{\Gamma(d/2)}} M_S^{2+d} R^d£¬
\end{equation}
where $M_P = 1/\sqrt{G_N} \sim 1.22\times10^{19}~GeV$ and $G_N$ is
Newton's constant. In our work we use the de Donder gauge for the
pure KK-graviton part and the Feynman gauge ($\xi=1$) for the SM
part. The Feynman rules for the relevant vertices including KK
graviton used in our calculations are given below \cite{13}, where
we assume that all the momenta flow to the vertices, except that the
fermionic momenta are set to be along the fermion flow directions.
\begin{itemize}
\item
$G_{\rm KK}^{\mu
\nu}(k_3)-\bar{\psi}(k_1)-\psi(k_2)~\textrm{vertex}: $
\begin{eqnarray}
-i \frac{\kappa}{8} \left[\gamma^{\mu} (k_1 + k_2)^{\nu} +
\gamma^{\nu} (k_1 + k_2)^{\mu} - 2 \eta^{\mu \nu} (\rlap/{k}_1 +
\rlap/{k}_2 - 2 m_{\psi}) \right],
\end{eqnarray}
\item
$G_{\rm KK}^{\mu \nu}(k_3)-Z^{\rho}(k_1)-Z^{\sigma}(k_2)~\textrm{vertex}: $
\begin{eqnarray}
i \kappa \left[B^{\mu \nu \rho \sigma} m_Z^2 + (C^{\mu \nu \rho
\sigma \tau \beta} - C^{\mu \nu \rho \beta \sigma \tau}) k_{1\tau}
k_{2\beta} + \frac{1}{\xi}E^{\mu \nu \rho \sigma}(k_1,k_2)\right],
\end{eqnarray}
\item
$G_{\rm KK}^{\mu
\nu}(k_4)-\bar{\psi}(k_1)-\psi(k_2)-Z^{\rho}(k_3)~\textrm{vertex}:
$
\begin{eqnarray}
i e \frac{\kappa}{4} \left( \gamma^{\mu} \eta^{\nu \rho} +
\gamma^{\nu} \eta^{\mu \rho} - 2 \gamma^{\rho}\eta^{\mu \nu}
\right)(v_f-a_f \gamma^5),
\end{eqnarray}
\end{itemize}
where $G_{\rm KK}^{\mu \nu}$, $\psi$ and $Z^{\mu}$ represent the
fields of graviton, lepton, and $Z^0$-boson, respectively,
$e=g\sin\theta_W$ is the positron electric charge, $\xi$ is the
$SU(2)$ gauge fixing parameter, $v_f$, $a_f$ are the vector and
axial-vector couplings which are the same as those defined in the
SM, and $\kappa = \sqrt{16 \pi G_N} = \sqrt{2}/\overline{M}_P$
where the reduced Planck scale $\overline{M}_P = M_P/\sqrt{8 \pi}$.
The explicit expressions for the tensor coefficients are given as
\begin{eqnarray}
B^{\mu \nu \alpha \beta} & = & \frac{1}{2}
      (\eta^{\mu \nu}\eta^{\alpha \beta}
      -\eta^{\mu \alpha}\eta^{\nu \beta}
      -\eta^{\mu \beta}\eta^{\nu \alpha}),
       \nb \\
C^{\rho \sigma \mu \nu \alpha \beta} & = & \frac{1}{2}
      [\eta^{\rho \sigma}\eta^{\mu \nu}\eta^{\alpha \beta}
     -(\eta^{\rho \mu}\eta^{\sigma \nu}\eta^{\alpha \beta}
      +\eta^{\rho \nu}\eta^{\sigma \mu}\eta^{\alpha \beta}
      +\eta^{\rho \alpha}\eta^{\sigma \beta}\eta^{\mu \nu}
      +\eta^{\rho \beta}\eta^{\sigma \alpha}\eta^{\mu \nu})],
      \nb \\
E^{\mu \nu \rho \sigma}(k_{1},k_{2}) & = &
      \eta^{\mu \nu}(k_1^{\rho} k_1^{\sigma} + k_2^{\rho} k_2^{\sigma}
      + k_1^{\rho} k_2^{\sigma}) - \left [\eta^{\nu \sigma} k_1^{\mu} k_1^{\rho}
      + \eta^{\nu \rho} k_2^{\mu} k_2^{\sigma} + (\mu \leftrightarrow \nu)\right ].
\end{eqnarray}
After summation over KK states the spin-2 KK graviton propagator can be expressed
as \cite{YJZhou}
\begin{eqnarray}
\tilde{G}_{\rm KK}^{\mu \nu \alpha \beta}=\frac{1}{2} D(s)
\left[\eta^{\mu \alpha} \eta^{\nu \beta} + \eta^{\mu \beta}
\eta^{\nu \alpha} - \frac{2}{d+2}\eta^{\mu \nu} \eta^{\alpha \beta}
\right],
\end{eqnarray}
where
\begin{equation}
D(s) = {{16\pi}\over{\kappa}^2}
{s^{d/2-1}\over{M_S}^{d+2}}
\biggl[\pi + 2i I(\Lambda/\sqrt{s})\biggr],
\end{equation}
and
\begin{equation}
I(\Lambda/\sqrt{s}) = P \int_0^{\Lambda/\sqrt{s}}dy\
{y^{d-1}\over 1-y^2}.
\end{equation}
The integral $I(\Lambda/\sqrt{s})$ should be understood that a point
$y=1$ has been removed from the integration path, and we set the
ultraviolet cutoff $\Lambda$ to be the fundamental scale $M_S$
routinely.

\par
\begin{figure}[!b]
  \centering
  \includegraphics{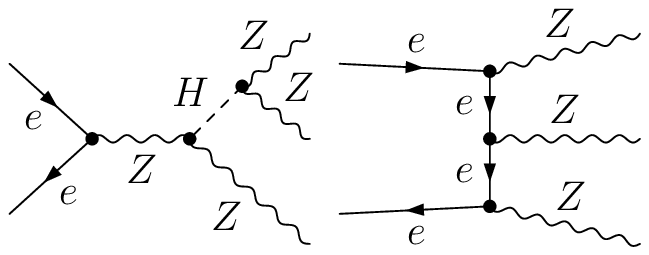}\\
  (a)\\
  \includegraphics{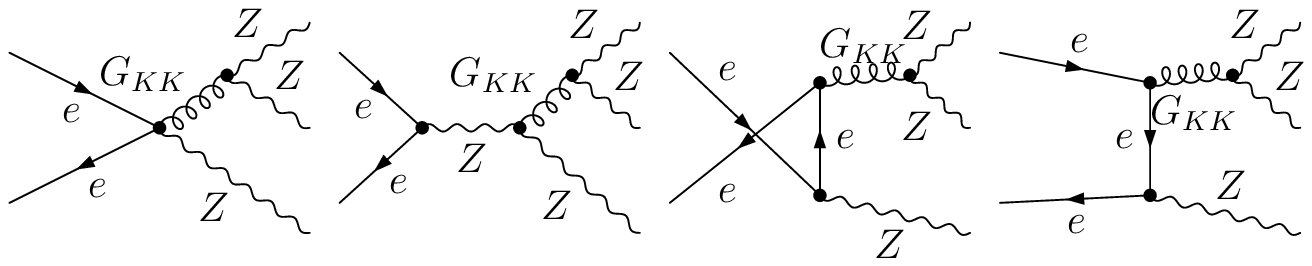}\\
  (b)\\
  \caption{The Feynman diagrams for the process \eezzz in the LED model.
  (a) The SM-like diagrams. (b) The additional diagrams with KK graviton exchange.}\label{fig1}
\end{figure}

We denote the process of the triple $Z^0$-boson production at the ILC as
\begin{equation}
e^+(p_1) + e^-(p_2) \to Z^0(p_3) + Z^0(p_4) + Z^0(p_5).
\end{equation}
The SM-like diagrams for the above process are depicted in
Fig.\ref{fig1}(a). In the LED model, the KK graviton can couple to
$Z^0$-pair and fermion-pair. We present the four additional pure LED
Feynman diagrams in Fig.\ref{fig1}(b).

\par
In our calculations the developed FeynArts3.4 package \cite{14} is
adopted to generate all the lowest order Feynman diagrams and
convert them to the corresponding amplitudes. Subsequently, the
amplitude calculations are mainly implemented by applying modified
FormCalc5.4 programs \cite{15}.

\vskip 5mm
\section{Numerical results and discussions}
\par
In our numerical calculations, we use the following set of input
parameters \cite{17}:
\begin{eqnarray}
& m_W = 80.385~GeV,~m_Z = 91.1876~GeV,~\alpha_{ew}(0) = 1/137.036, & \nb \\
& \sin^{2}\theta_W = 1 - m_W^2/m_Z^2 = 0.222897,~m_e =
0.510998928~MeV. & \nb
\end{eqnarray}
We know that the ATLAS and CMS experiments found several SM
Higgs-like events at the location of $M_H \sim 125~ GeV$
\cite{17-1}. Recently, ATLAS reported the searching for extra
dimensions by using diphoton events in $\sqrt{s} = 7~TeV$ $pp$
collisions \cite{18}. The results provided $95\%$ C.L. lower limits
on the fundamental Planck scale $M_S$ between $2.27~TeV$ and
$3.53~TeV$ depending on the number of extra dimensions $d$ in the
range of $7$ to $3$. The diphoton and dilepton results from CMS set
limits on $M_S$ in the range of $2.5 - 3.8~TeV$ as $d$ varies from
$7$ to $2$ at $95\%$ C.L. \cite{19}. In this work we take
$M_H=125~GeV$, and set $M_S = 3.5~TeV$ (or $M_S= 3.8~TeV$) and $d =
3$ as the representative ADD parameters in case otherwise stated.

\par
In Refs.\cite{11,12} there exist the calculations for the SM
one-loop electroweak corrections to the \eezzz process. We make
comparison of our LO numerical results with theirs, there we adopt
the input parameters equal to those in Ref.\cite{11} and use two
different packages in order to check the correctness of our LO
calculations. In Table \ref{tab1} we present the results of the LO
integral cross sections in the SM at the $\sqrt{s}=500~GeV$ ILC
obtained by using CompHEP-4.5.1 \cite{16} and
FeynArts3.4/FormCalc5.4 packages separately, and the LO
$\sigma_{SM}$ results by using FeynArts3.3/FormCalc5.3 package
presented in Table 1 of Ref.\cite{11}. We can see that all the
corresponding cross sections are in good agreement within the
calculation errors.
\begin{table}[!tbp]
  \centering
  \begin{tabular}{|c|c|c|c|c|c|c|}
  \hline
  {}$M_H$      &\multicolumn{3}{c|}{$\sigma_{\rm SM}~(fb)$ at LO}         \\ \cline{2-4}
  {}$(GeV)$         &CompHEP        &FeynArts       & Ref.\cite{11}      \\
  \hline    115     &1.0053(3)      &1.0055(1)      &1.0055(2)      \\
  \hline    120     &1.0136(3)      &1.0138(1)      &1.0138(2)      \\
  \hline    150     &1.0974(3)      &1.0974(1)      &1.0975(2)      \\
  \hline    170     &1.2563(4)      &1.2564(1)      &1.2564(2)      \\
  \hline
\end{tabular}
\caption{The numerical results of the LO integral cross sections for the
process \eezzz in the SM at the $\sqrt{s}=500~GeV$ ILC by using CompHEP-4.5.1
and FeynArts3.4/FormCalc5.3 packages, and the LO $\sigma_{SM}$ by using
FeynArts3.3/FormCalc5.3 package provided in Table 1 of Ref.\cite{11}. }\label{tab1}
\end{table}

\par
\begin{figure}[!htbp]
  \centering
  \includegraphics[scale=0.7]{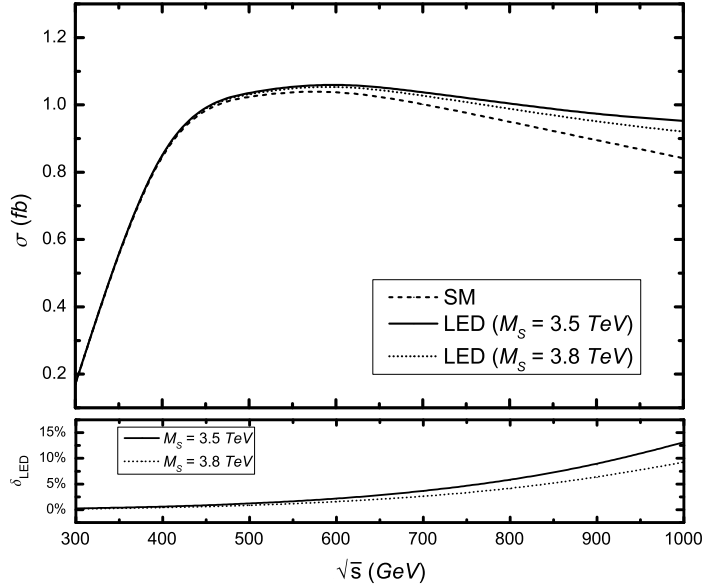}\\
\caption{The integrated cross sections for the process \eezzz in
both the SM and the LED model, and the relative discrepancy due to
the LED effect $\left(\delta_{LED}\equiv
\frac{\sigma_{LED}-\sigma_{SM}}{\sigma_{SM}}\right)$, as functions
of the colliding energy $\sqrt{s}$ by taking $M_S = 3.5~TeV$,
$3.8~TeV$ and $d = 3$.}\label{fig2}
\end{figure}

In the upper plot of Fig.\ref{fig2} we present the numerical results
of the LO integrated cross sections for the process \eezzz as
functions of the colliding energy $\sqrt{s}$ in both the SM and the
LED model. The curves reveal that the integrated cross section
increases rapidly when $\sqrt{s} < 500~GeV$ and decreases smoothly
when $\sqrt{s} > 600~GeV$. Obviously, we can find that the curve for
the LED model decreases more slowly than that for the SM in the
region of $\sqrt{s} > 600~GeV$. We define the relative discrepancy
as $\delta_{LED}\equiv \frac{\sigma_{LED}-\sigma_{SM}}{\sigma_{SM}}$
to describe the LED effect on the integral cross section for the
process \eezzz, and plot its distribution versus $\sqrt{s}$ in the
nether plot of Fig.\ref{fig2}. We can read out from Fig.\ref{fig2}
that the LED relative discrepancies at the positions of
$\sqrt{s}=500~GeV$, $800~TeV$, and $1~TeV$ are $1.15\%$ $(0.83\%)$,
$5.69\%$ $(4.07\%)$, and $13.11\%$ $(9.27\%)$ for $M_S =
3.5~(3.8)~TeV$, separately. The LED relative discrepancy
$\delta_{LED}$ goes up as the increment of the colliding energy
$\sqrt{s}$, and the enhancement of the cross section
$\sigma_{LED}-\sigma_{SM}$ becomes larger and larger when the
colliding energy $\sqrt{s}$ goes beyond $600~GeV$.

\par
\begin{figure}[!htbp]
  \centering
  \includegraphics[scale=0.52,bb= 25 10 450 320]{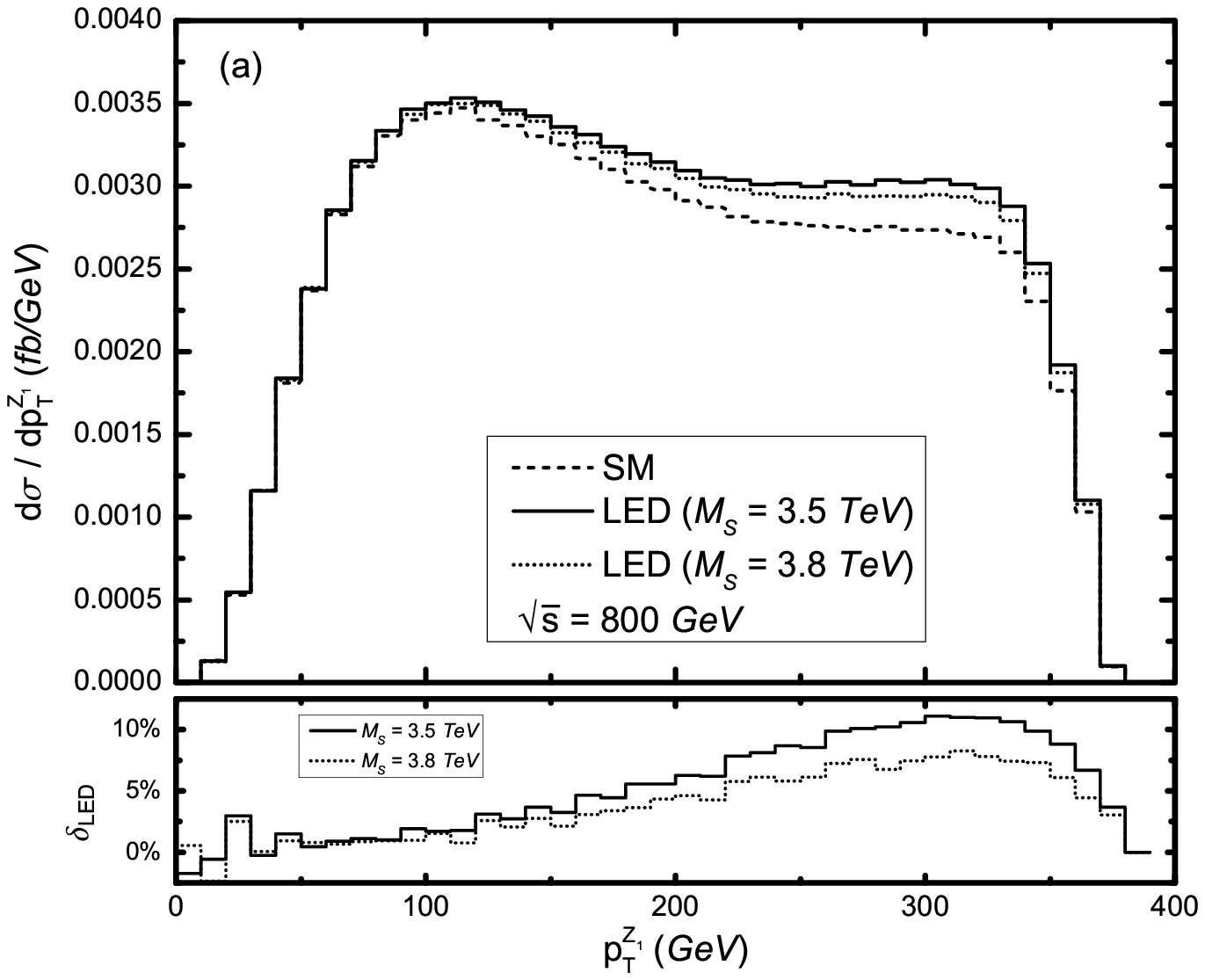}%
  \hspace{0in}%
  \includegraphics[scale=0.52,bb= 25 10 450 320]{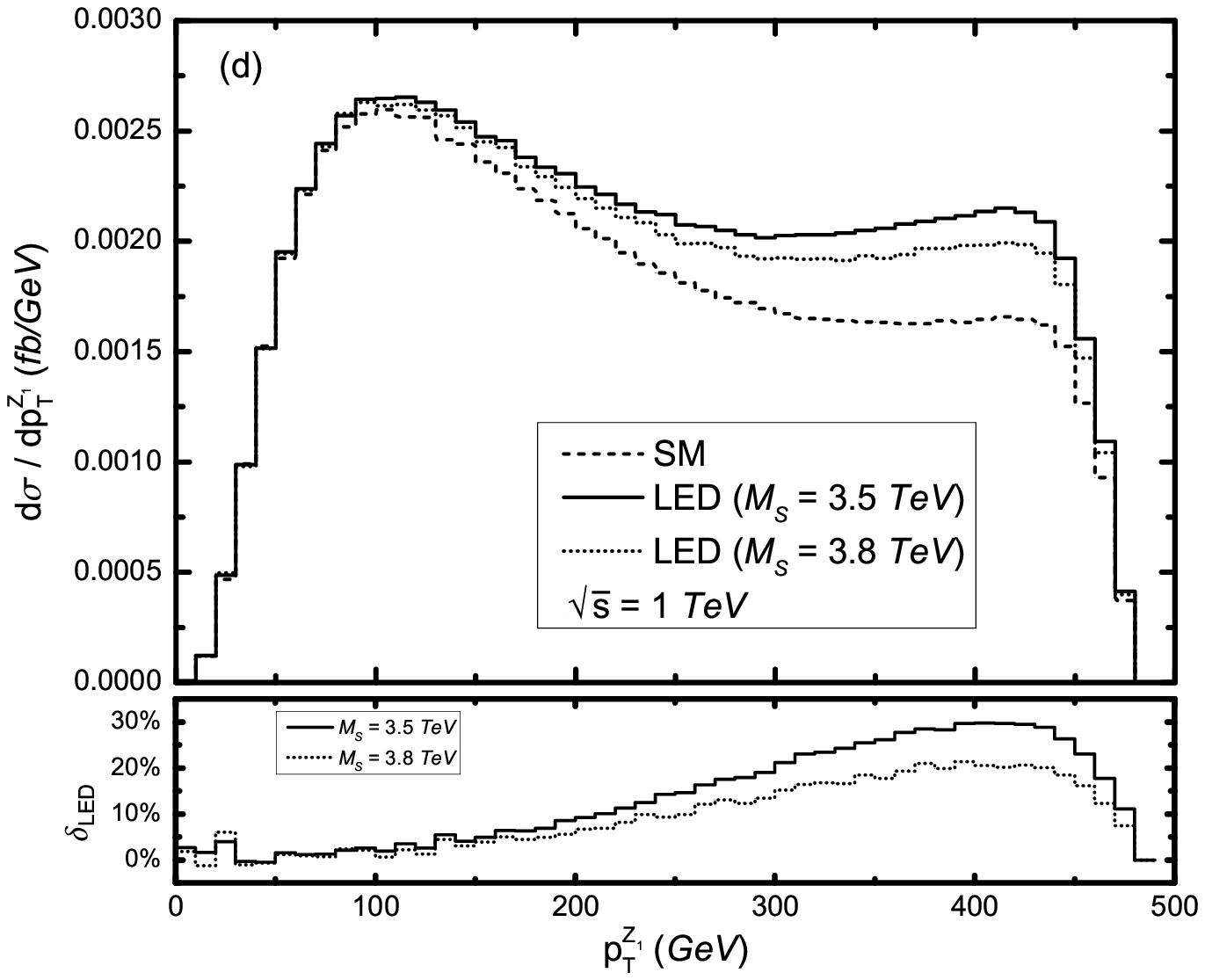}%
  \hspace{0in}% \\
  (a)~~~~~~~~~~~~~~~~~~~~~~~~~~~~~~~~~~~~~~~~~~~~~~~~~~~~~~~~(d) \\[2mm]
  \includegraphics[scale=0.52,bb= 25 10 450 320]{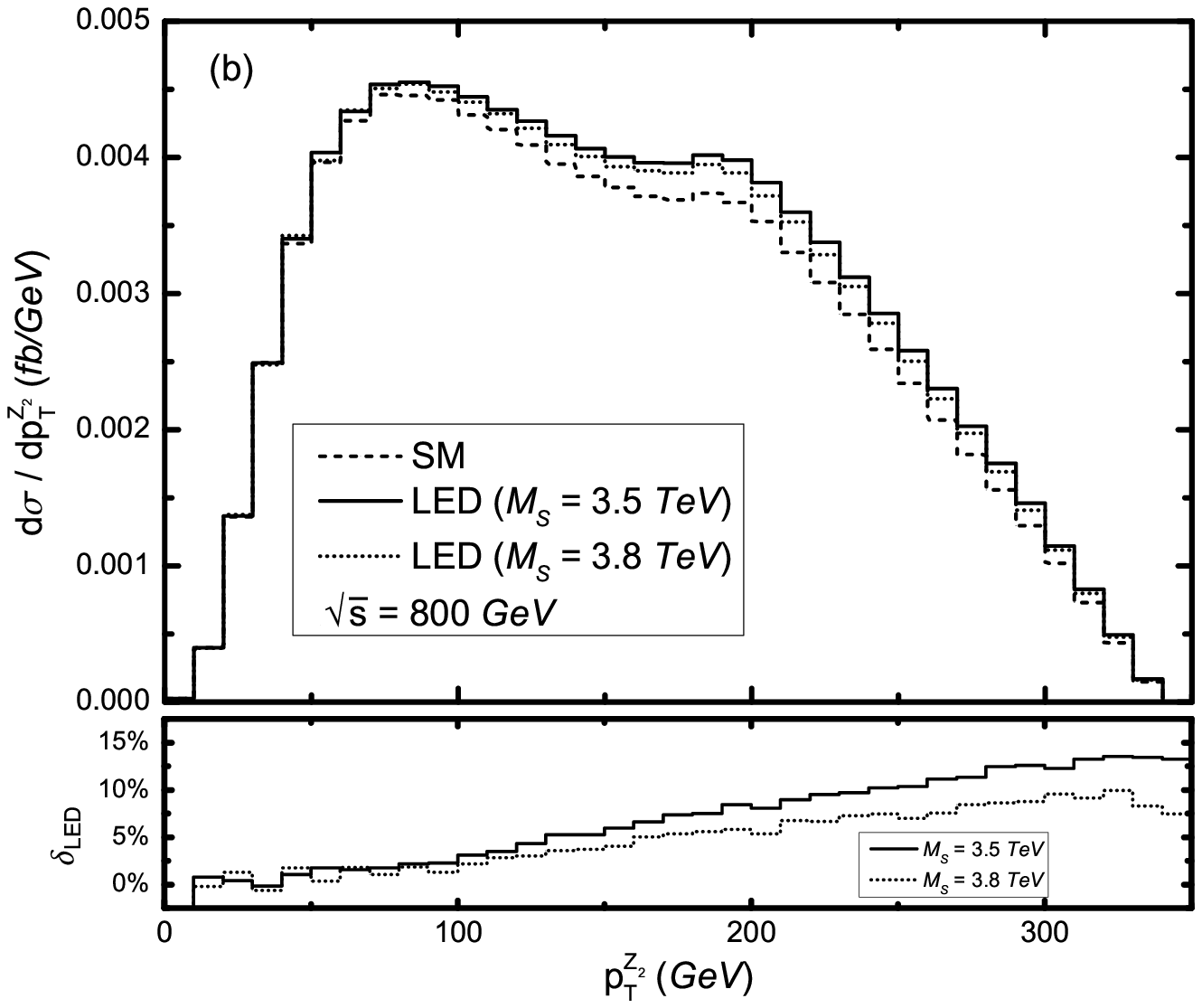}%
  \hspace{0in}%
  \includegraphics[scale=0.52,bb= 25 10 450 320]{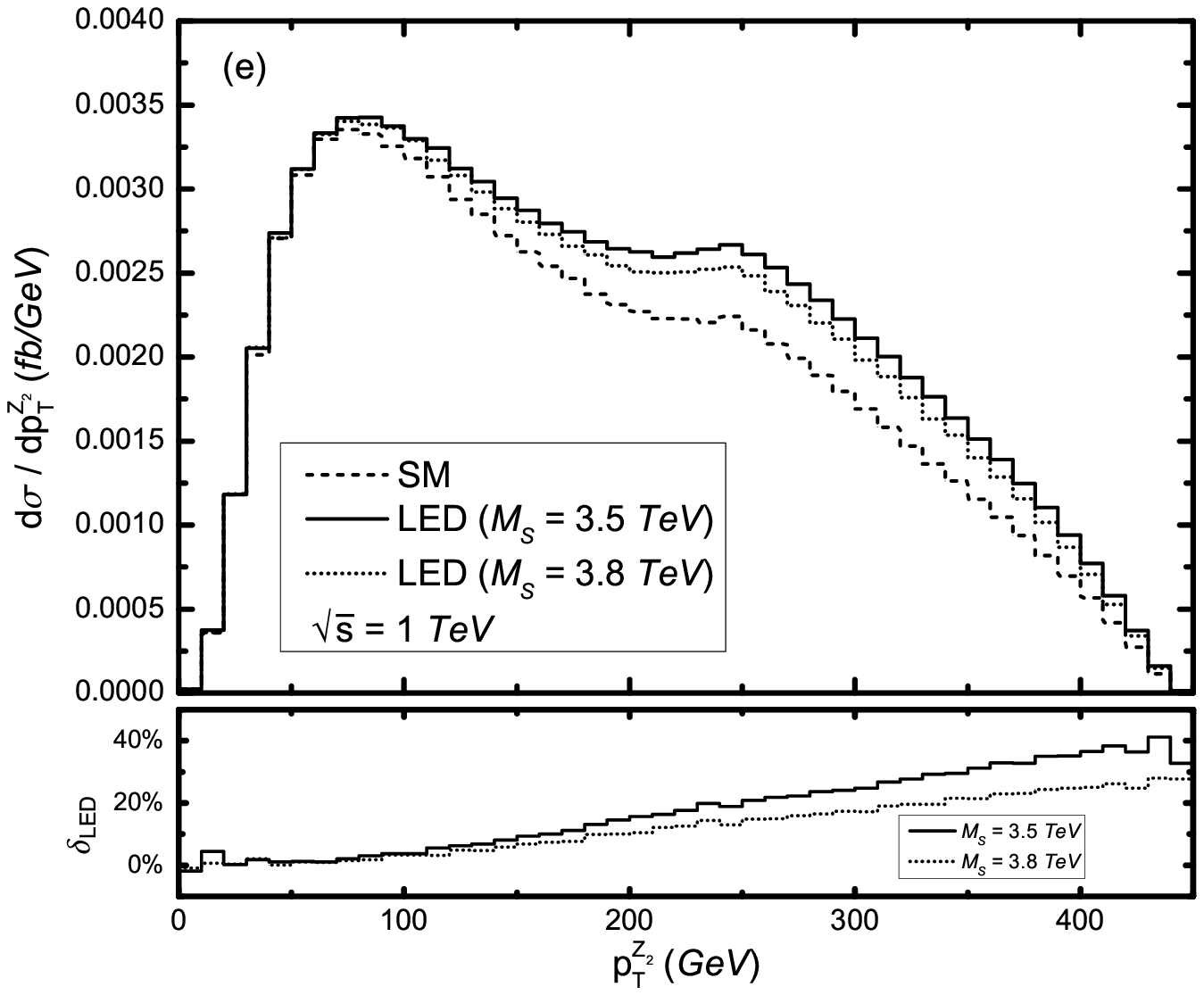}%
  \hspace{0in}% \\
  (b)~~~~~~~~~~~~~~~~~~~~~~~~~~~~~~~~~~~~~~~~~~~~~~~~~~~~~~~~(e) \\[2mm]
  \includegraphics[scale=0.52,bb= 25 10 450 320]{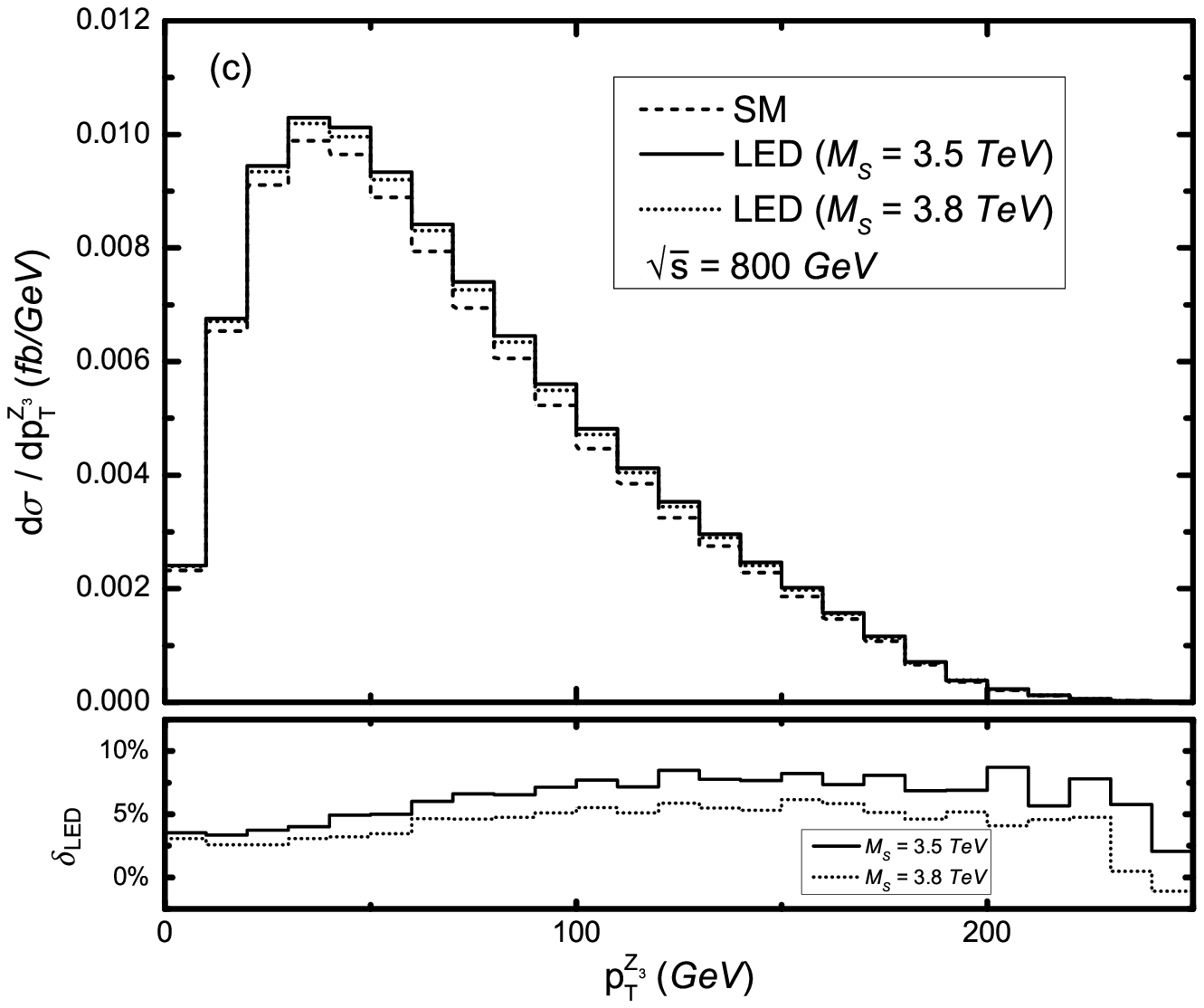}%
  \hspace{0in}%
  \includegraphics[scale=0.52,bb= 25 10 450 320]{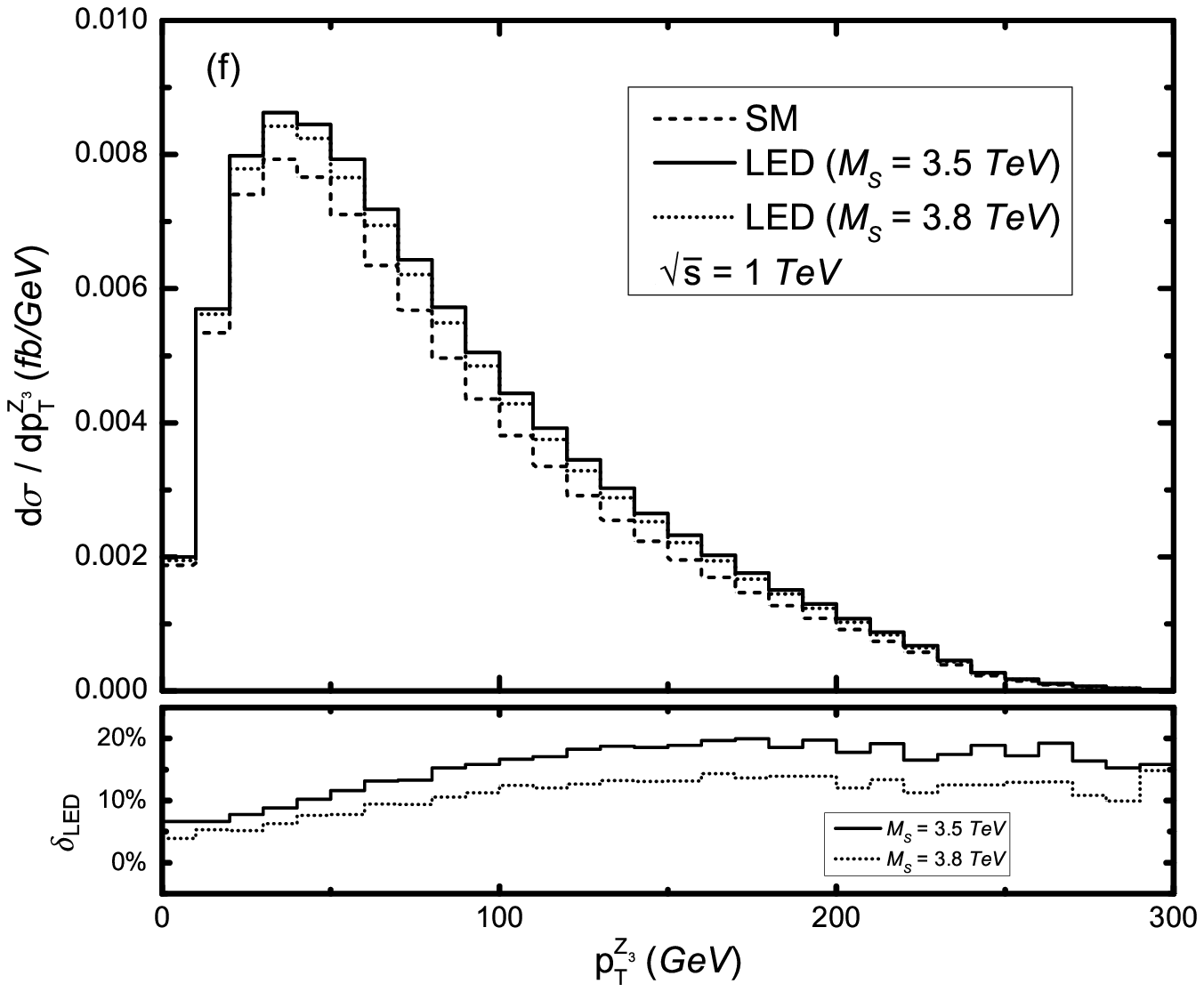}%
  \hspace{0in}% \\
  (c)~~~~~~~~~~~~~~~~~~~~~~~~~~~~~~~~~~~~~~~~~~~~~~~~~~~~~~~~(f)
\caption{\label{fig3} The distributions of the transverse momenta of
$Z^0$-bosons in the SM and the LED model, and the corresponding
relative discrepancies, defined as $\delta_{LED}(p_T^Z) \equiv
\left(\frac{d\sigma_{LED}}{dp_T^Z}-\frac{d\sigma_{SM}}{dp_T^Z}\right)/
\frac{d\sigma_{SM}}{dp_T^Z}$, with $M_S = 3.5,~3.8~TeV$ and $d = 3$.
Plots (a), (b) and (c) are for the $p_{T}^{Z_1}$,  $p_{T}^{Z_2}$ and
$p_{T}^{Z_3}$ distributions at the $\sqrt{s} = 800~ GeV$ ILC, while
(d), (e) and (f) are for the $p_T^{Z_{1,2,3}}$ distributions at the
$\sqrt{s} = 1~ TeV$ ILC, respectively. }
\end{figure}

In analyzing the \eezzz event, we classify the final three $Z^0$
bosons as the leading $Z^0$-boson, the next-to-leading $Z^0$-boson
and the next-to-next-to-leading $Z^0$-boson according to their
transverse momenta, labeled as $Z_1$, $Z_2$ and $Z_3$, respectively.
The criterion for final $Z^0$-boson clarification is based on the
conditions of $p_T^{Z_1} > p_T^{Z_2} > p_T^{Z_3}$. In Fig.\ref{fig3}
we present the distributions of the transverse momenta of
$Z_{1,2,3}$-bosons in the SM and the LED model, and the
corresponding relative discrepancies,
$\frac{d\sigma_{SM}}{dp_T^{Z_{1,2,3}}}$,
$\frac{d\sigma_{LED}}{dp_T^{Z_{1,2,3}}}$,
$\delta_{LED}(p_T^{Z_{1,2,3}})$ in the conditions of $M_S =
3.5~TeV,~3.8~TeV$, $\sqrt{s} = 800~GeV,~1~TeV$, separately. The
curves in Fig.\ref{fig3} show that the LED effect enhances the
differential cross section, particularly in the relative high
transverse momentum region. We can see that in the conditions of
$\sqrt{s} = 800~GeV$ and $M_S = 3.5~(3.8)~TeV$, the relative
discrepancies $\delta_{LED}(p_T^{Z_{1,2,3}})$ can reach the maximal
values of $11.08\%$ $(8.27\%)$, $13.54\%$ $(9.92\%)$ and $8.72\%$
$(6.17\%)$ separately, while in the conditions of $\sqrt{s}=1~TeV$
and $M_S = 3.5~(3.8)~TeV$, the maximum values of
$\delta_{LED}(p_T^{Z_{1,2,3}})$ are increased to be $29.75\%$ $(21.39\%)$,
$41.14\%$ $(27.96\%)$, and $19.94\%$ $(14.35\%)$, respectively.
The curves for the differential cross sections of $p_T^{Z_{1}}$ in Figs.\ref{fig3}(a,b)
demonstrate the maxima values of $\delta_{LED}(p_T^{Z_{1}})$ are located
at the vicinities of $p_T^{Z_{1}} = 310~GeV$ and $p_T^{Z_{1}} = 410~GeV$ at the
$\sqrt{s} = 800~GeV$ and $1~TeV$ ILC respectively, while
Fig.7 in Ref.\cite{7} shows the relative discrepancy increases
monotonously with the increment of $p_T^{Z_{1}}$ at the LHC.
\begin{figure}[!htbp]
  \centering
  \includegraphics[scale=0.52,bb= 25 10 450 320]{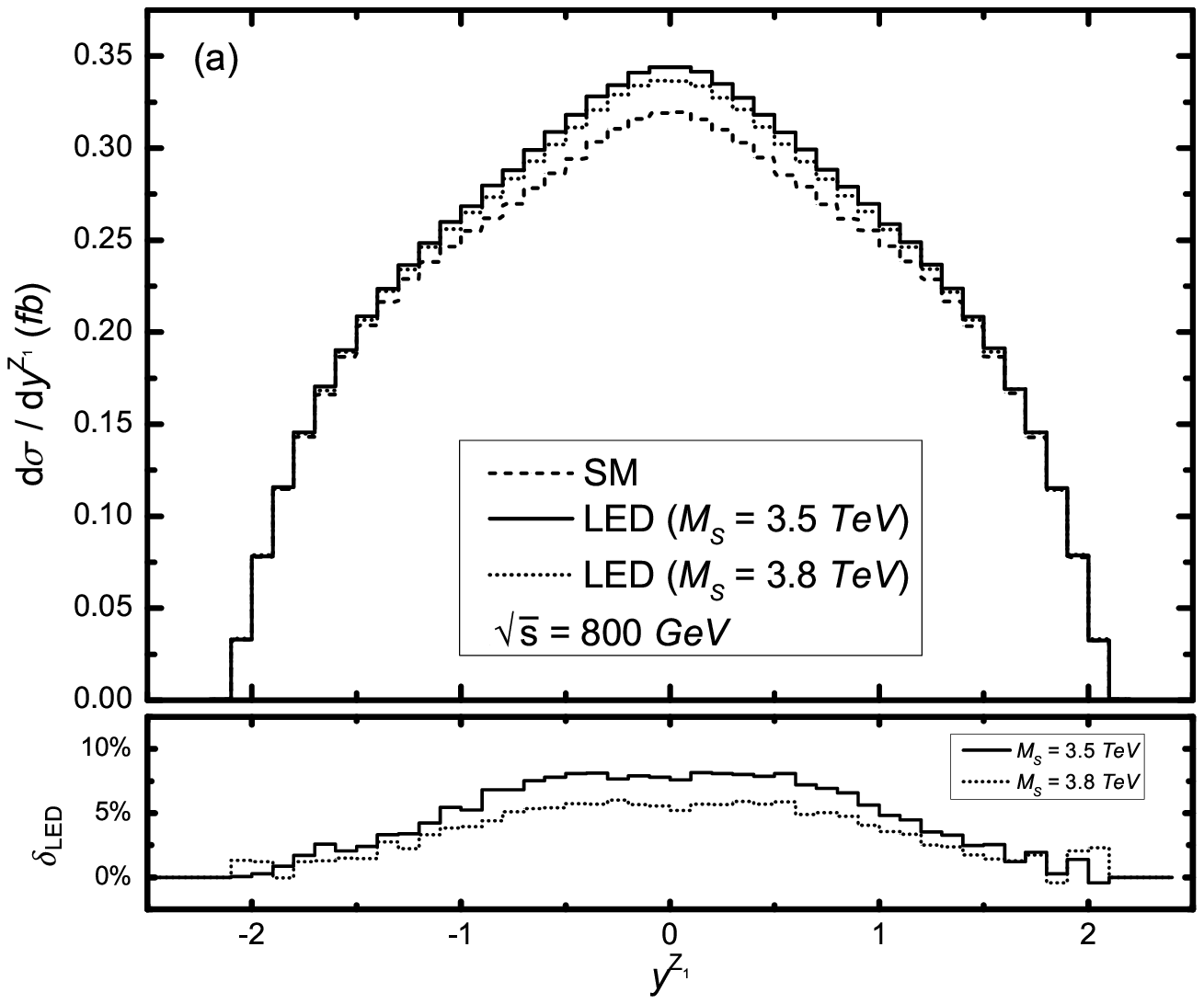}%
  \hspace{0in}%
  \includegraphics[scale=0.52,bb= 25 10 450 320]{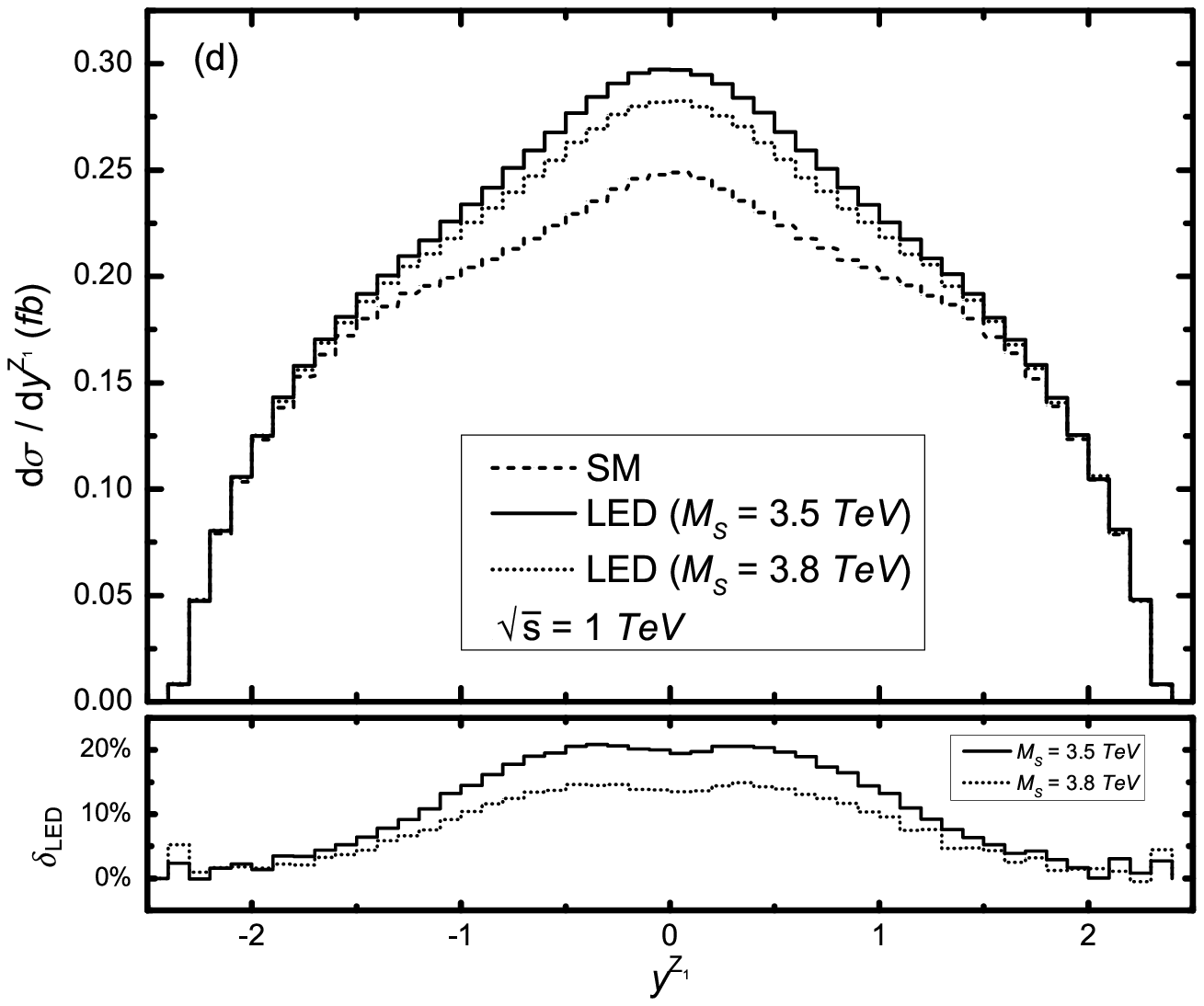}%
  \hspace{0in}% \\
  (a)~~~~~~~~~~~~~~~~~~~~~~~~~~~~~~~~~~~~~~~~~~~~~~~~~~~~~~~~(d) \\[2mm]
  \includegraphics[scale=0.52,bb= 25 10 450 320]{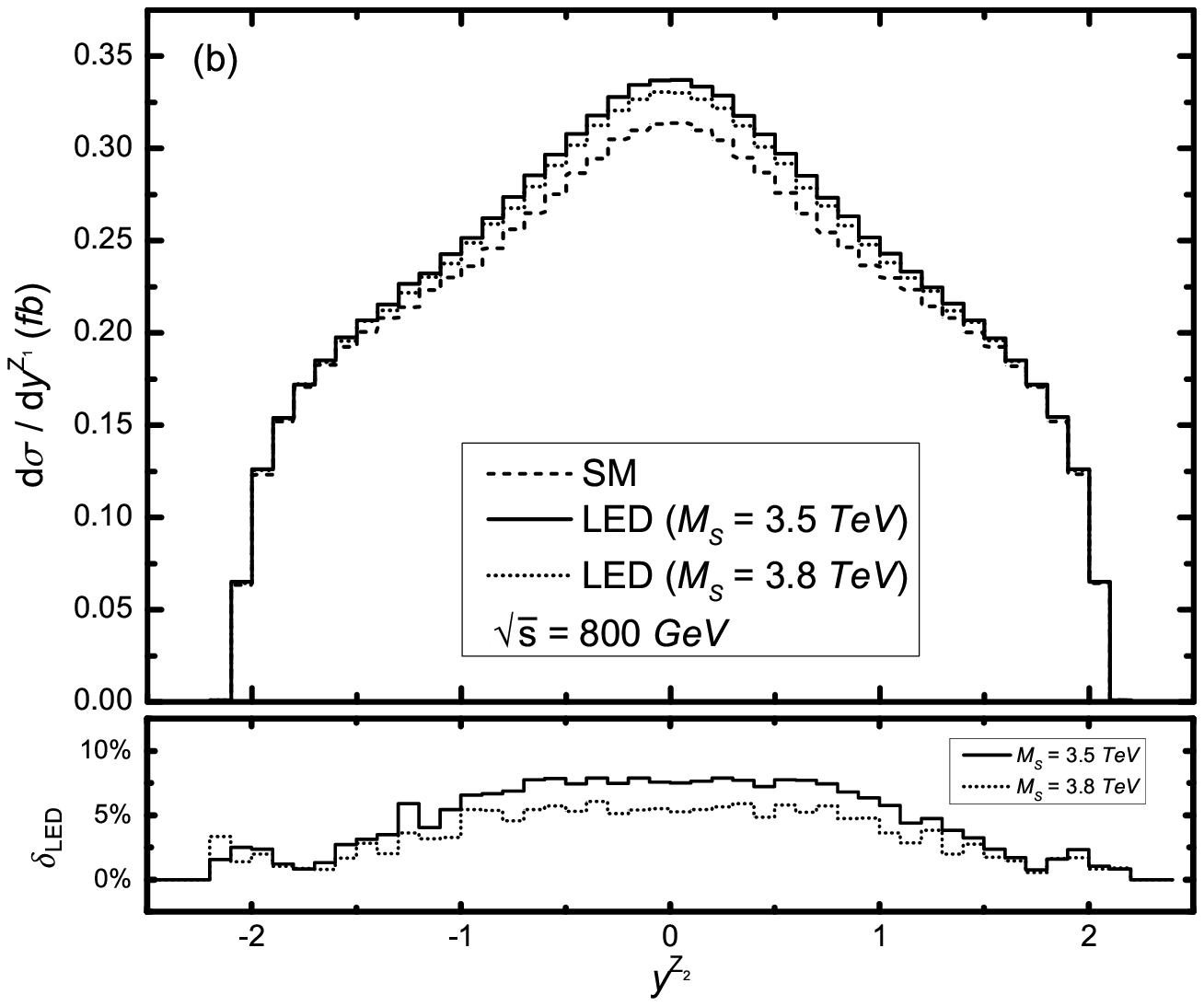}%
  \hspace{0in}%
  \includegraphics[scale=0.52,bb= 25 10 450 320]{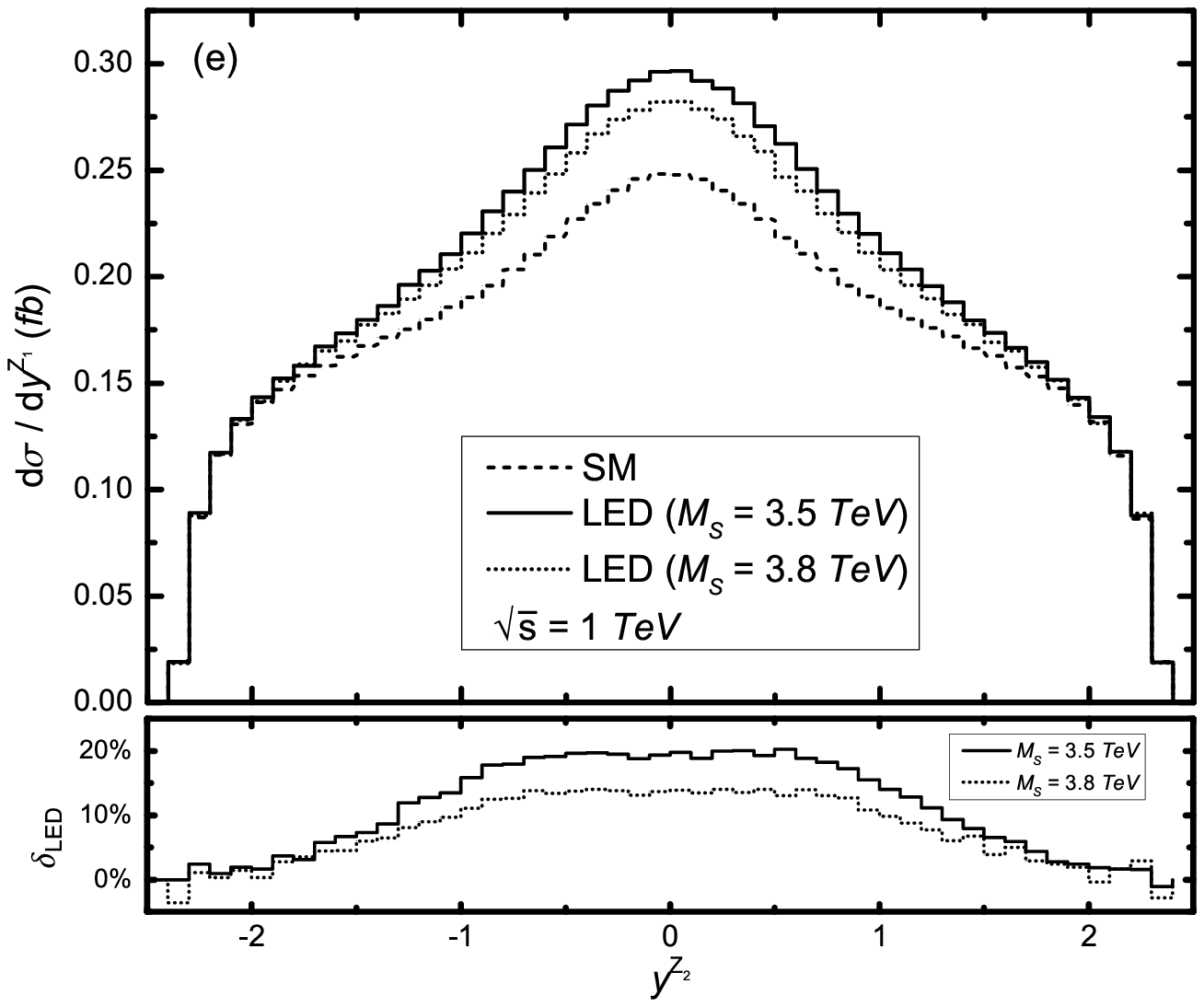}%
  \hspace{0in}% \\
  (b)~~~~~~~~~~~~~~~~~~~~~~~~~~~~~~~~~~~~~~~~~~~~~~~~~~~~~~~~(e) \\[2mm]
  \includegraphics[scale=0.52,bb= 25 10 450 320]{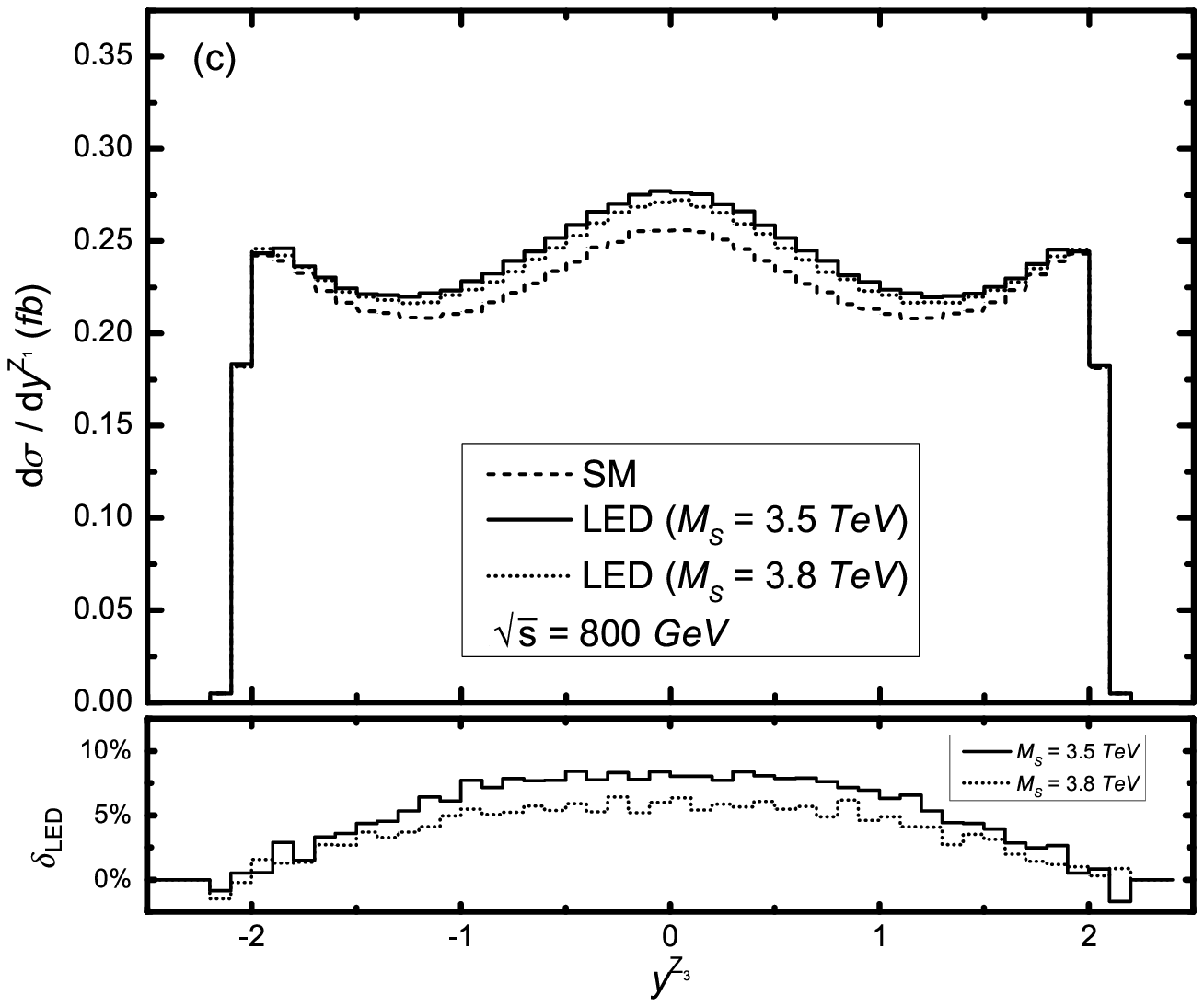}%
  \hspace{0in}%
  \includegraphics[scale=0.52,bb= 25 10 450 320]{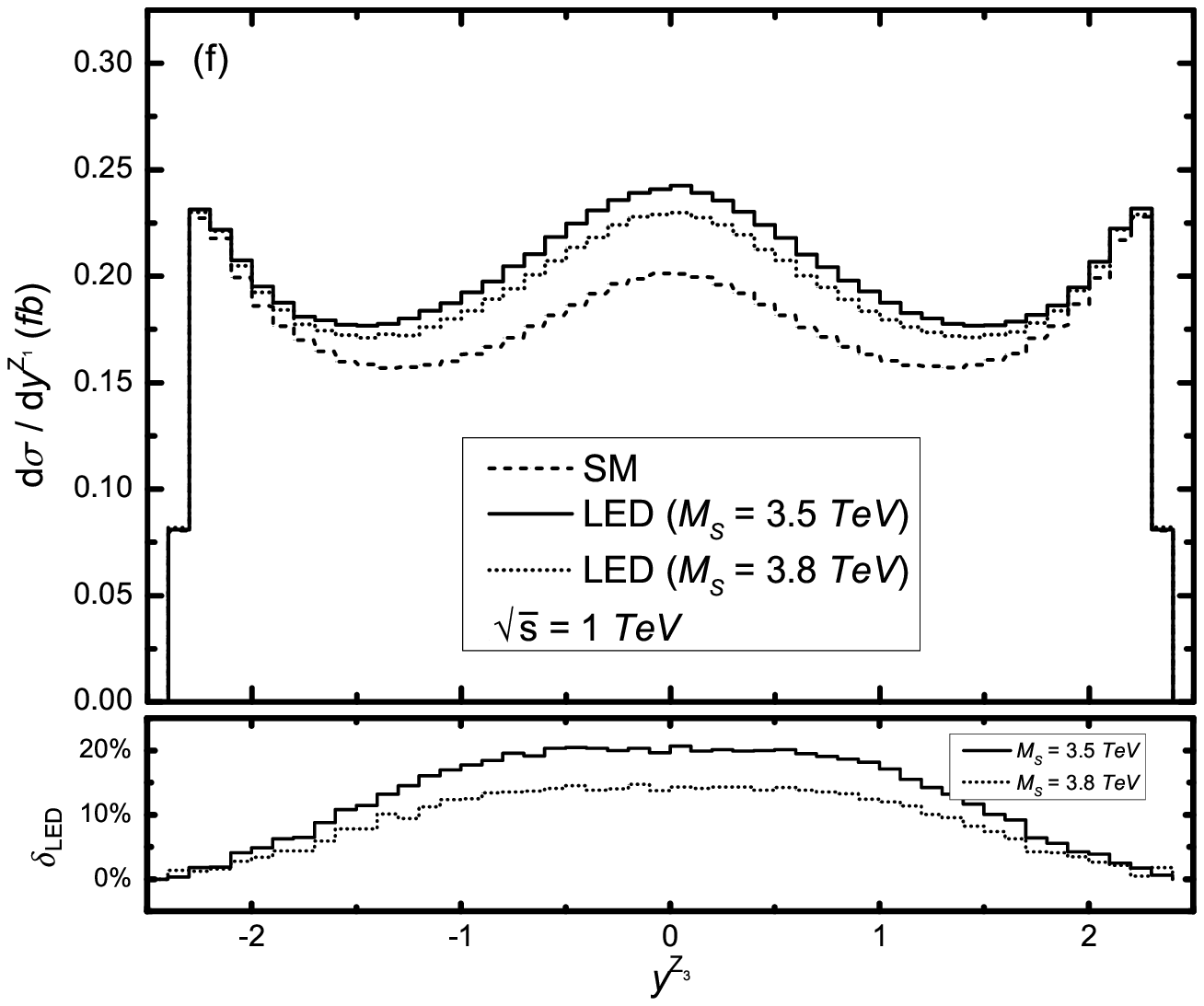}%
  \hspace{0in}% \\
  (c)~~~~~~~~~~~~~~~~~~~~~~~~~~~~~~~~~~~~~~~~~~~~~~~~~~~~~~~~(f)
  \caption{The rapidity distributions of $Z^0$-bosons in the SM and the LED model,
and the corresponding relative discrepancies, defined as
$\delta_{LED}(y^Z) \equiv
\left(\frac{d\sigma_{LED}}{dy^Z}-\frac{d\sigma_{SM}}{dy^Z}\right)/\frac{d\sigma_{SM}}{dy^Z}$,
with $M_S = 3.5,~3.8~TeV$ and $d = 3$. Figs.\ref{fig4}(a), (b) and
(c) are for the $y_{T}^{Z_1}$, $y_{T}^{Z_2}$ and  $y_{T}^{Z_3}$
distributions at the $\sqrt{s} = 800~ GeV$ ILC, while (d), (e) and
(f) are for the $y_T^{Z_{1,2,3}}$ distributions at the $\sqrt{s} =
1~ TeV$ ILC, respectively.  }\label{fig4}
\end{figure}

\par
The rapidity distributions of final $Z$-bosons in the SM and the LED
model, $d \sigma_{SM, LED}/dy^{Z_{1,2,3}}$, and the corresponding
LED relative discrepancies, $\delta_{LED}(y^{Z_{1,2,3}})$, with $M_S
= 3.5~TeV,~3.8~TeV$ and $\sqrt{s} = 800~GeV,~1~TeV$ are shown in
Figs.\ref{fig4}(a-f), respectively. The line-shapes of $y^{Z_1}$ and
$y^{Z_2}$ distributions are similar, while the $y^{Z_3}$
distribution seems to be particular. All the curves for specific
$Z^0$-boson, such as $Z_1$ (or $Z_2$, $Z_3$), with different values
of $M_S$ and $\sqrt{s}$ look to be generally similar. The plots in
the Fig.\ref{fig4} show that the curves bulge apparently in the
central rapidity region, and the relative discrepancies with  $M_S =
3.5~(3.8)~TeV$ can reach about $8\%$ $(6\%)$ and $20\%$ $(14\%$) in
the central rapidity regions at the $\sqrt{s} = 800~GeV$ and
$\sqrt{s} = 1~TeV$ ILC, separately. The LED effects intensify the
cross section evidently in the central rapidity region of $|y|<0.8$
where $\delta_{LED}(y^{Z_{1,2,3}})$ are all beyond $5\%$, and the
relative discrepancies are basically stable in this region.

\par
\begin{figure}[!htbp]
  \centering
  \includegraphics[scale=0.52,bb= 25 20 450 320]{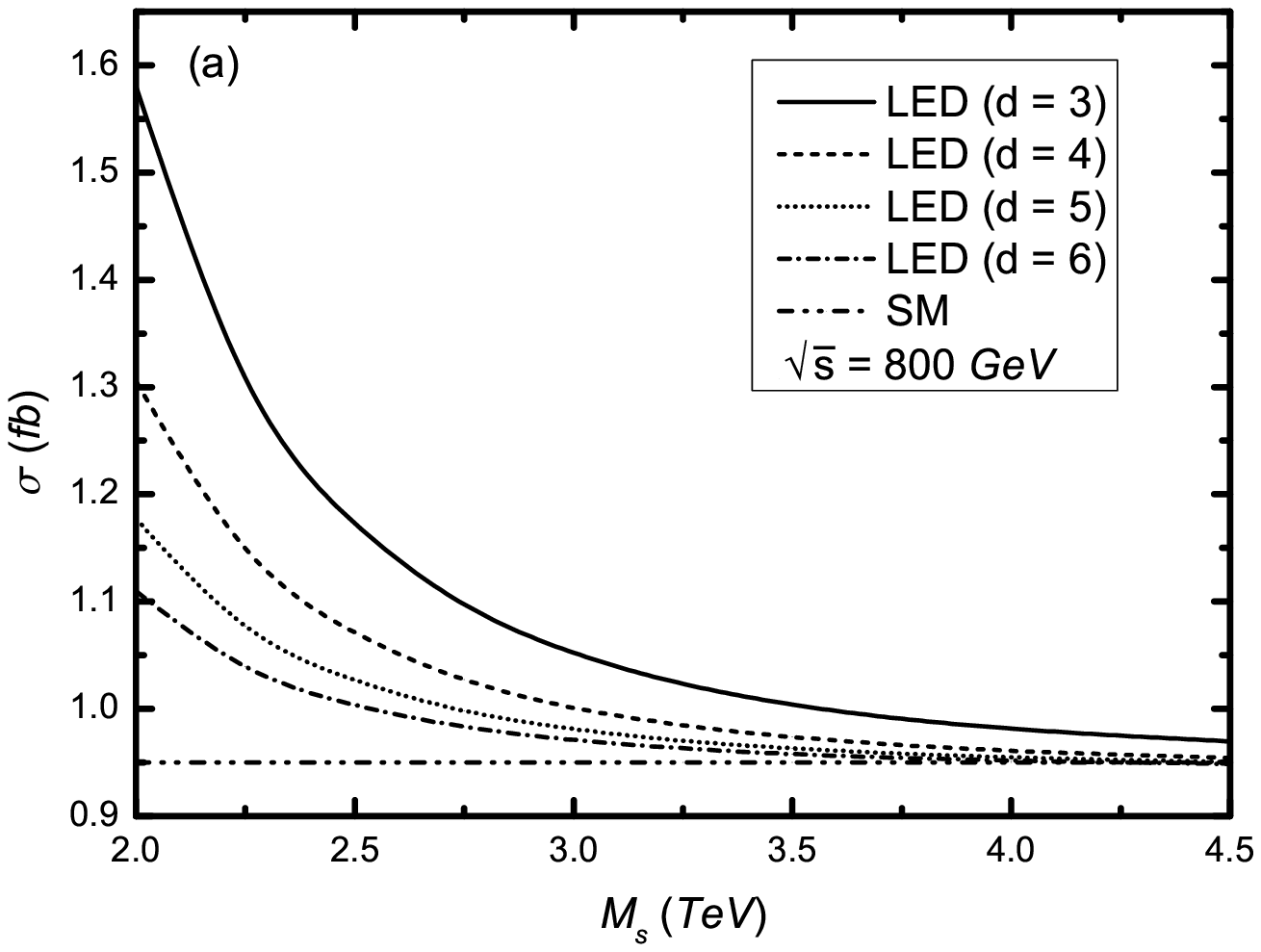}%
  \hspace{0in}%
  \includegraphics[scale=0.52,bb= 25 20 450 320]{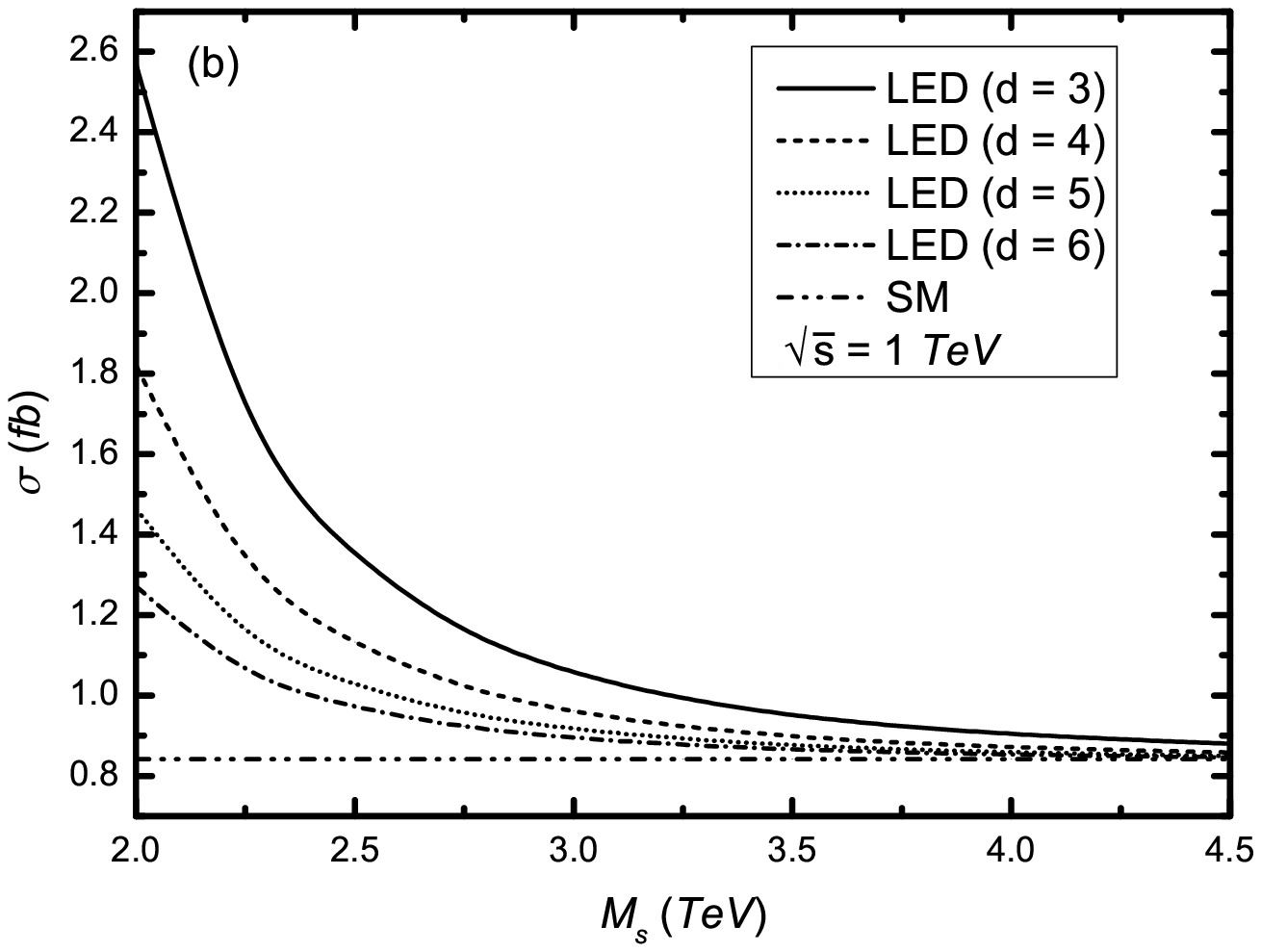}%
  \hspace{0in}%
  (a)~~~~~~~~~~~~~~~~~~~~~~~~~~~~~~~~~~~~~~~~~~~~~~~~~~~~~~~~(b)
  \caption{The cross sections in the LED model as functions of $M_S$ with
  $d=3,4,5,6$. The additional horizontal line is for the SM cross sections. (a)
  $\sqrt{s} = 800~GeV$. (b) $\sqrt{s} = 1~TeV$.}\label{fig5}
\end{figure}

In Figs.\ref{fig5}(a,b) we present the cross sections in the LED model as
functions of $M_S$ for different $d$ values at the $\sqrt{s}=800~GeV$ and
$\sqrt{s}=800~GeV$ ILC, separately. The SM cross section
which is independent of these two LED parameters is depicted as a
horizontal line. Obviously, Fig.\ref{fig5} shows that as the
increment of either $M_S$ or $d$, the cross section in the LED model
obviously decreases and are getting close to the SM prediction.

\par
\begin{figure}[!htbp]
  \centering
  \includegraphics[scale=0.52,bb= 25 10 450 320]{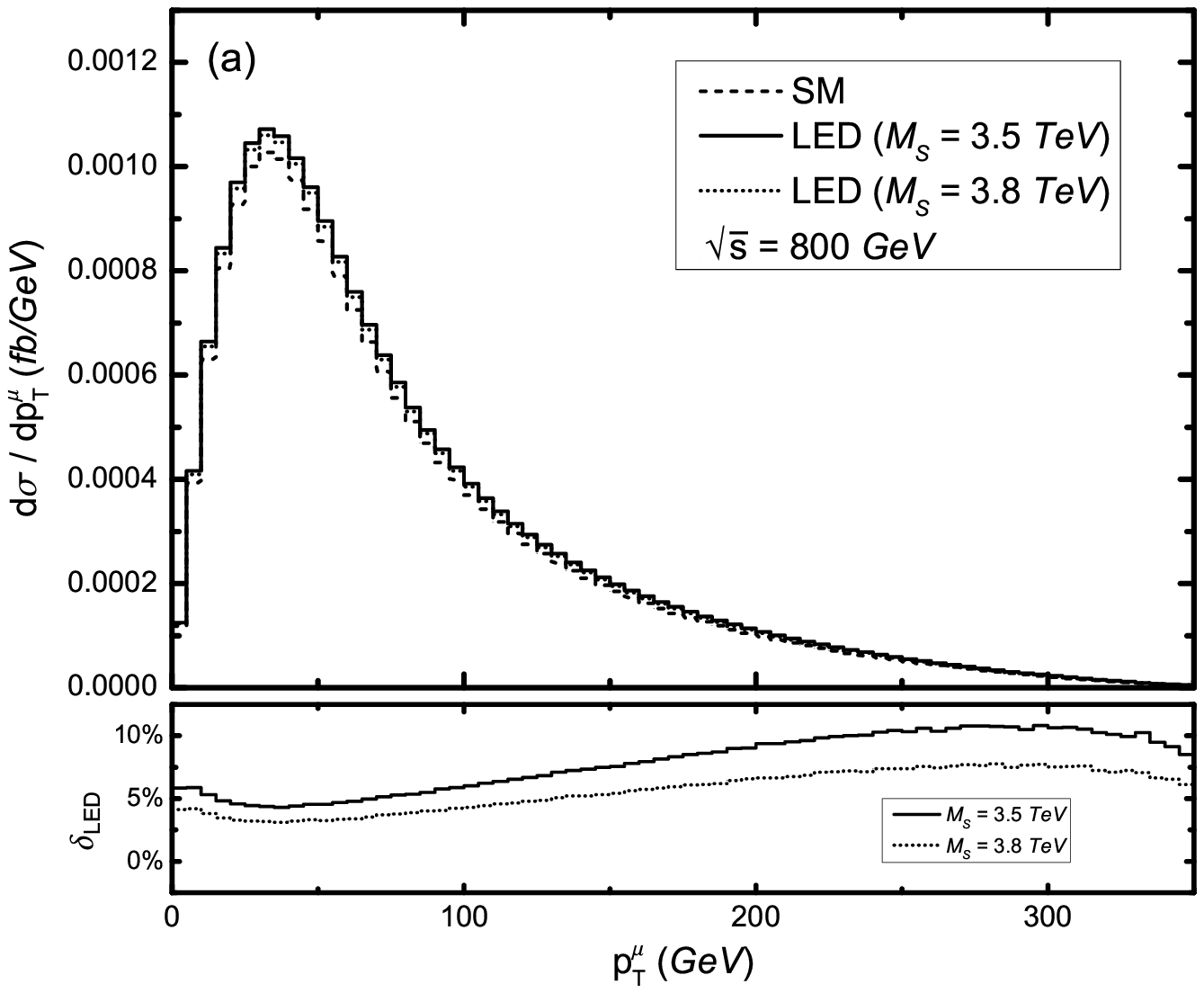}%
  \hspace{0in}%
  \includegraphics[scale=0.52,bb= 25 10 450 320]{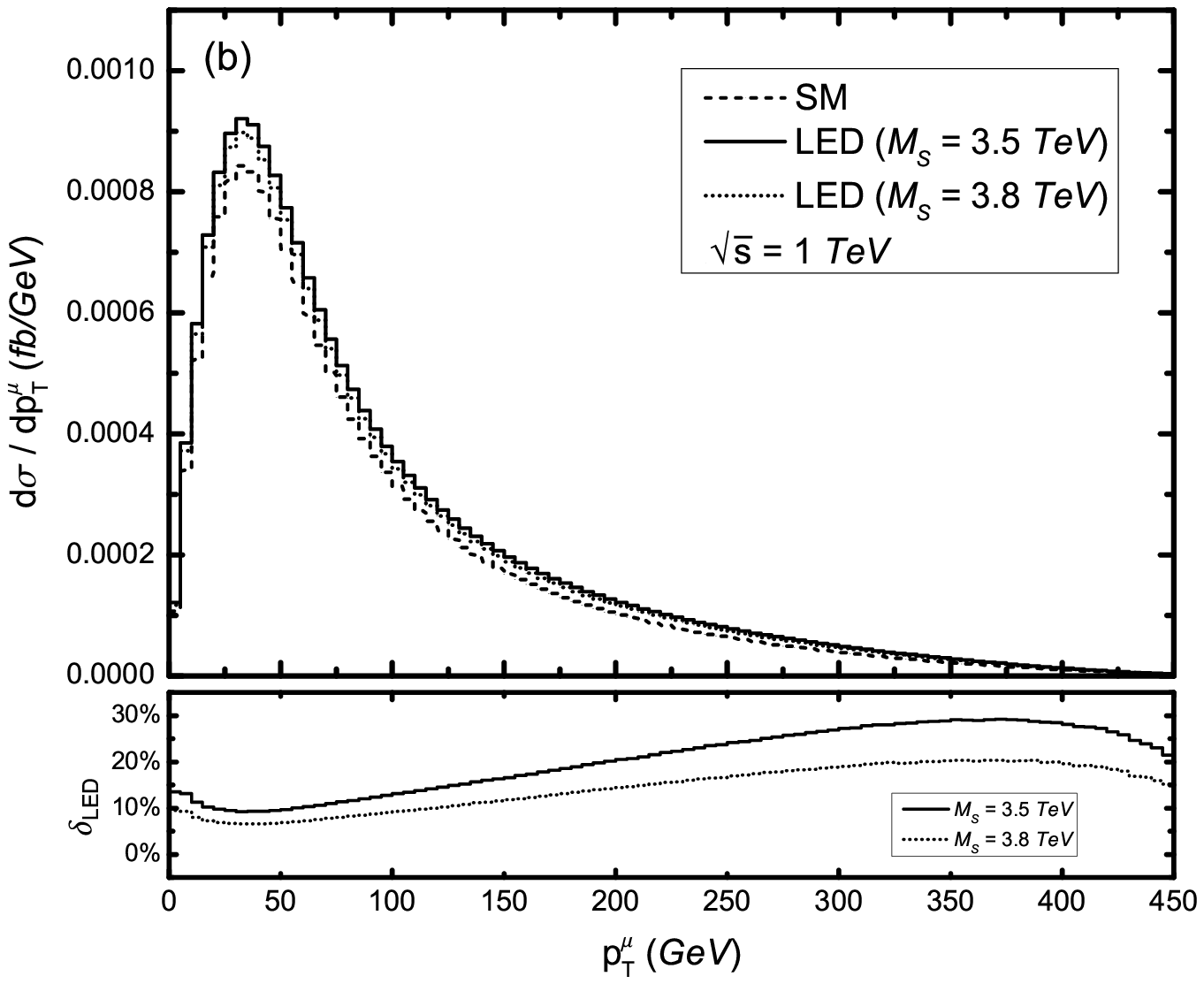}%
  \hspace{0in}% \\
  (a)~~~~~~~~~~~~~~~~~~~~~~~~~~~~~~~~~~~~~~~~~~~~~~~~~~~~~~~~(b) \\[2mm]
\caption{The distributions of the transverse momentum of final muon in the
SM and the LED model, and the corresponding relative discrepancies
(defined as $\delta_{LED}(p_T^{\mu}) \equiv (\frac{d\sigma_{LED}}{dp_T^{\mu}}
- \frac{d\sigma_{SM}}{dp_T^{\mu}}) / \frac{d\sigma_{SM}}{dp_T^{\mu}}$) with $M_S
= 3.5~TeV,~3.8~TeV$ and $d = 3$. (a) at the $\sqrt{s}$ = 800 $GeV$ ILC. (b) at the
$\sqrt{s} = 1~TeV$ ILC.}\label{fig6}
\end{figure}

In the following we consider the inclusive process of
\begin{equation}
e^+ e^- \to Z^0 Z^0 Z^0 \to \mu^+ \mu^- +X,
\end{equation}
where the muons are produced by the subsequential $Z^0$ decay of $Z^0 \to \mu^+\mu^-$.
We take the branch ratio as $Br(Z^0 \to \mu^+ \mu^-)
= 3.366\%$ \cite{19}. Since the kinematic distributions
of final $\mu^+$ and $\mu^-$ are the same, we present the distributions of the
transverse momentum of $\mu$ in both the SM and the LED model, $\frac{d\sigma_{SM}}{dp_T^{\mu}}$ and
$\frac{d\sigma_{LED}}{dp_T^{\mu}}$, and the corresponding relative discrepancies,
$\delta_{LED}(p_T^{\mu})$, with $d = 3$, $M_S = 3.5~TeV,~3.8~TeV$ and
$\sqrt{s} = 800~GeV,~1~TeV$ in Figs.\ref{fig6}(a) and (b), respectively. We can read out from the
figures that the $\delta_{LED}(p_T^{\mu})$ can reach the maximal
values of $10.83\%$ and $7.78\%$ for $M_S = 3.5~TeV$ and $M_S=3.8~TeV$,
respectively when $\sqrt{s} = 800~GeV$, while in the case of
$\sqrt{s} = 1~TeV$ the maximum values of $\delta_{LED}(p_T^{\mu})$
increase to $29.16\%$ and $20.37\%$ for $M_S = 3.5~TeV$
$M_S=3.8~TeV$, respectively.

\par
\begin{figure}[!htbp]
  \centering
  \includegraphics[scale=0.52,bb= 25 10 450 320]{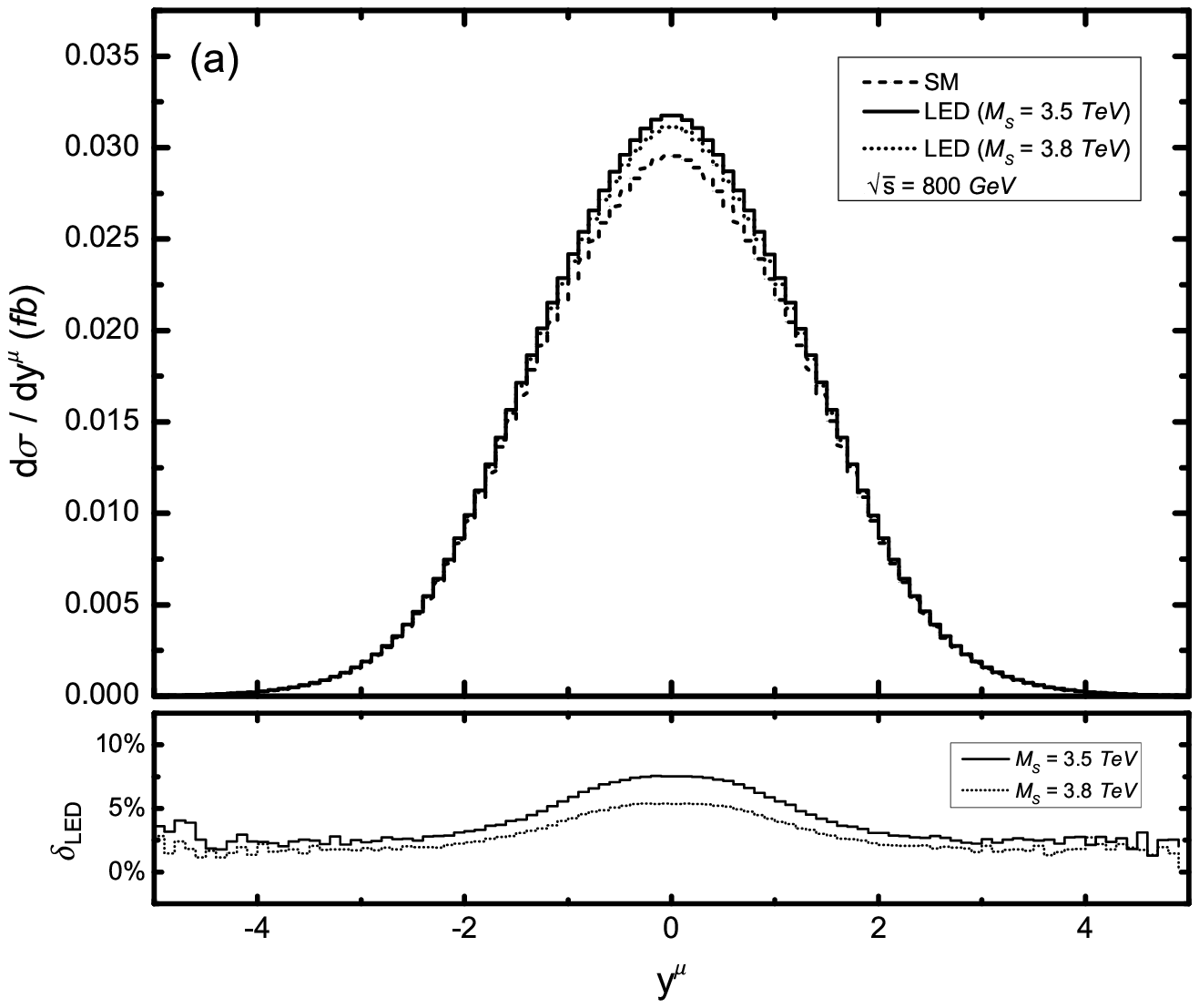}%
  \hspace{0in}%
  \includegraphics[scale=0.52,bb= 25 10 450 320]{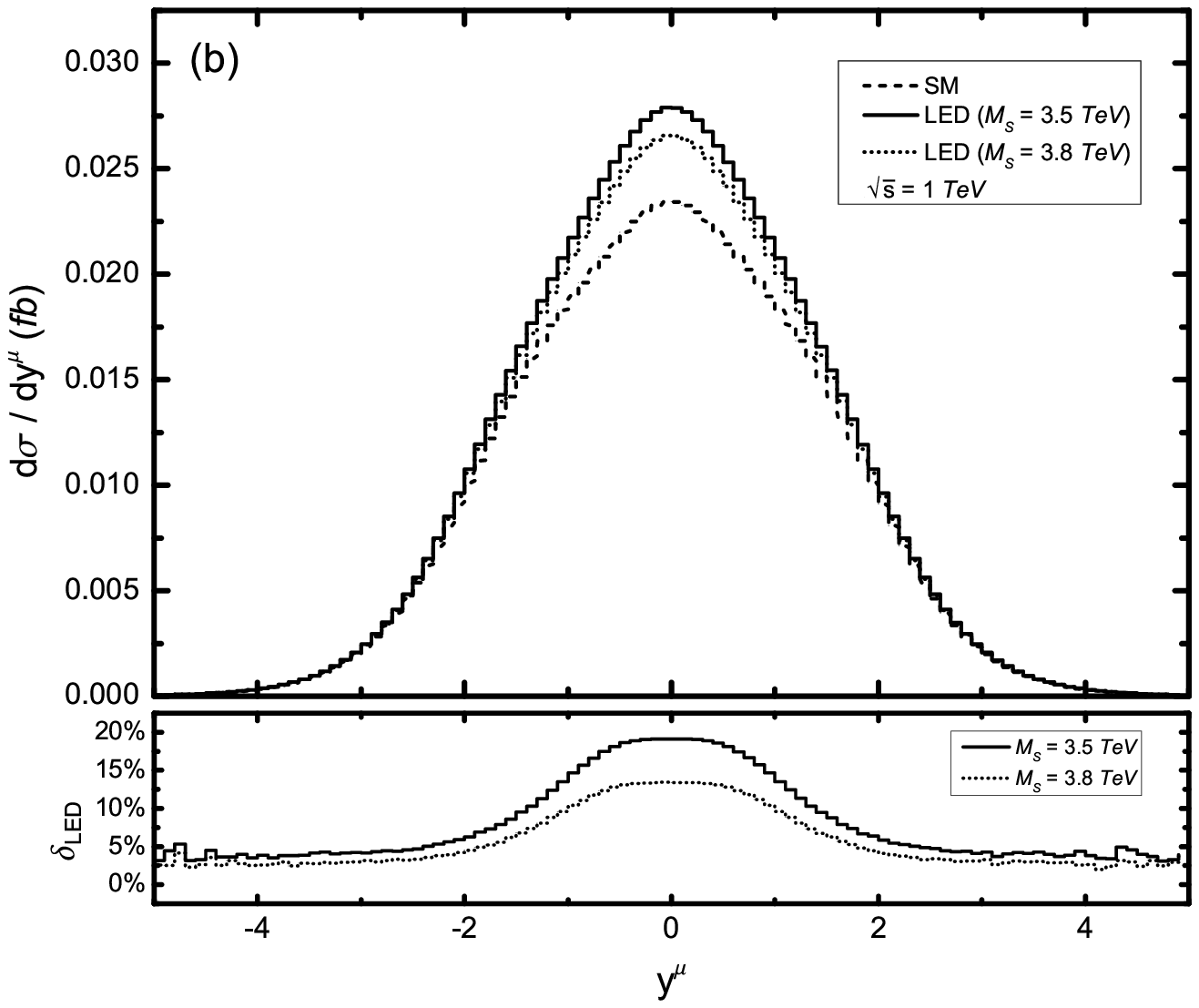}%
  \hspace{0in}% \\
  (a)~~~~~~~~~~~~~~~~~~~~~~~~~~~~~~~~~~~~~~~~~~~~~~~~~~~~~~~~(d) \\[2mm]
\caption{The rapidity distributions of muon in the SM and the LED model,
and the corresponding relative discrepancies (defined as $\delta_{LED}(y^{\mu})
\equiv (\frac{d\sigma_{LED}}{dy^{\mu}} - \frac{d\sigma_{SM}}{dy^{\mu}}) /
\frac{d\sigma_{LED}}{dy^{\mu}}$) with $M_S = 3.5~TeV,~3.8~TeV$ and $d=3$.
(a) at the $\sqrt{s} = 800~GeV$ ILC. (b) at the $\sqrt{s} = 1~TeV$ ILC.}\label{fig7}
\end{figure}

The rapidity distributions of final muon in the SM and the LED
model, $\frac{d\sigma_{SM}}{dy^{\mu}}$ and $\frac{d\sigma_{LED}}{dy^{\mu}}$,
and the corresponding LED relative discrepancies,
$\delta_{LED}(y^{\mu})$, with $d = 3$, $M_S = 3.5~TeV,~3.8~TeV$ and
$\sqrt{s}= 800~GeV,~1~TeV$ are shown in Figs.\ref{fig7}(a,b),
separately. The curves for the rapidity and the relative
discrepancy distributions in the central rapidity regions bulge apparently, and
the relative discrepancies have obvious peaks. There $\delta_{LED}(y^{\mu})$ for
$M_S = 3.5~TeV~(3.8~TeV)$ can reach $7.5\%~(5.4\%)$ and $19.1\%~(13.4\%)$ at the
$\sqrt{s} = 800~GeV$ and $\sqrt{s} = 1~TeV$ ILC, respectively.

\vskip 5mm
\section{Summary}
\par
In this paper we study the impact of KK graviton exchange on the
scattering process \eezzz in the LED model at the ILC. This process
is very useful in measuring the quartic gauge-boson couplings and
probing the existence of extra dimensions. We present the dependence
of the cross sections in both the SM and the LED model on the
colliding energy $\sqrt{s}$, and the kinematic distributions of
final $Z^0$ bosons and their subsequential decay products (muons) at
the ILC. We find the contribution from the KK graviton exchange
enhances the cross section evidently when $\sqrt{s}$ goes up beyond
$700~GeV$. We provide also the distributions of the transverse
momentum and rapidity of the final produced muon. The
results show that the LED effects become more observable in high
$p_T$ ranges or the central region of $|y| < 0.8$. We also
demonstrate the relationship between the cross sections in the LED
model and the fundamental Planck scale $M_S$ with the extra
dimensions $d$ are $3$, $4$, $5$, $6$, respectively.

\vskip 5mm
\par
\noindent{\large\bf Acknowledgments:} This work was supported in
part by the National Natural Science Foundation of China (Contract
No.11075150, No.11005101, No.11275190), and the Specialized Research Fund for the
Doctoral Program of Higher Education (Contract No.20093402110030).

\end{document}